\documentclass[preprint,prb,aps,superscriptaddress]{revtex4-1}
\usepackage{graphicx}
\usepackage{dcolumn}
\usepackage{bm}
\usepackage{lipsum}
\usepackage{ragged2e}
\usepackage{braket}
\usepackage{bbm}
\usepackage{color}
\usepackage{xcolor}
\usepackage{amsfonts}
\usepackage{amsmath}
\usepackage{dsfont}
\usepackage{natbib}
\providecommand{\keywords}[1]{\textbf{\textit{Index terms---}} #1}
\usepackage[mathlines]{lineno}
\usepackage[colorinlistoftodos]{todonotes}
\usepackage[colorlinks=true, allcolors=blue]{hyperref}

\begin{document}

\newcommand{\unamone}{Departamento de Sistemas Complejos, Instituto de Fisica,
Universidad Nacional Aut\'onoma de M\'exico, Apartado Postal 20-364,01000,
Ciudad de M\'exico, M\'exico.}
\newcommand{\unamtwo}{Instituto de F\'isica,
Universidad Nacional Aut\'onoma de M\'exico, Apartado Postal 20-364 01000,
Ciudad de  M\'{e}xico, M\'exico}
\newcommand{\uabc}{Facultad de Ciencias, Universidad Aut\'onoma de Baja California, Apartado Postal 1880, 22800 Ensenada, Baja California, M\'exico}

\title{Electronic spectrum of Kekul\'e patterned graphene considering second neighbor-interactions}

\author{Elías Andrade}
\affiliation{\unamone}
\author{Gerardo G. Naumis}
\affiliation{\unamone}
\author{R. Carrillo-Bastos}
\affiliation{\uabc}

\date{\today}

\begin{abstract}

The effects of second-neighbor interactions in Kekulé patterned graphene electronic properties are studied starting from a tight-binding Hamiltonian.  
Thereafter, a low-energy effective Hamiltonian is obtained by projecting the high energy bands at the $\Gamma$ point into the subspace defined by the Kekul\'e wave vector. The spectrum of the low energy Hamiltonian is in excellent agreement with the one obtained from a numerical diagonalization of the full tight-binding Hamiltonian. The main effect of the second-neighbour interaction is that a set of bands gains an effective mass and a shift in energy, thus lifting the degeneracy of the conduction bands at the Dirac point. This band structure is akin to a “spin-one Dirac cone”, a result expected for honeycomb lattices with a distinction between one third of the atoms in one sublattice. Finally, we present a study of Kekulé patterned graphene nanoribbons. This shows that the previous effects are enhanced as the width decreases. Moreover, edge states become dispersive, as expected due to second neighbors interaction, but here the Kek-Y bond texture results in an hybridization of both edge states. The present study shows the importance of second neighbors in realistic models of Kekulé patterned graphene, specially at surfaces.

\end{abstract}

\maketitle

%-----------------------------------------------------------------------
% Cover page
%-----------------------------------------------------------------------

%-----------------------------------------------------------------------
% Content page
%-----------------------------------------------------------------------

\setlength{\parskip}{1em}

\section{Intoduction}

The space-modulation of two-dimensional materials 
has opened avenues for new exciting physical phenomena
and applications \cite{MOGERA2020470,Wu_2020,NaumisReview,Chen2019,Yankowitz2018,Ni2015,Taboada2017,Ohta2012,Bistritzer,Zheng2016}. Several mechanisms allow to perform such modulations, these include interactions with substrates\cite{zhou2007substrate} as in  moire patterns\cite{ponomarenko2013cloning}, strain\cite{vozmediano2010gauge,amorim2016novel,naumis2017electronic}, adatoms\cite{Bianchi2010, Kaasbjerg2019}, magnetic fields\cite{novoselov2004electric,jiang2007infrared,guinea2006electronic}, and time dependent electromagnetic fields\cite{eckardt2015high,LaserGap,FloquetTop}.

Among such modulated systems, in graphene it has been experimentally observed that vacancies in a Cu substrate induce a spatial frequency  modulation with the size of an hexagonal ring of carbon atoms \cite{Gutierrez2016}. This modulated system is known as  kekul\'e-distorted graphene  \cite{Gutierrez2016,Gamayun,Elias_2019,Wu2020,Tijerina_2019,Hoi_2019,Penglin_2019}. As an example of its interest and importance, such kekul\'e distortion has been proposed as a possible mechanism behind superconductivity in magic-angle twisted bilayer graphene \cite{Bitan_2010,Hoi_2019}. Also, strain in kekul\'e distorted graphene can be used to perform valleytronics \cite{Elias_2019}, a feature that recently has been experimentally confirmed \cite{Daejin}.  Multiflavor Dirac fermions were predicted to emerge in kekul\'e graphene bilayers \cite{Tijerina_2019}, and it is even possible to produce such modulation in non-atomic systems, as with mechanical waves  in solids \cite{Mendez2020} and acoustical lattices \cite{Penglin_2019}.  Kekul\'e modulations are also reacheable via photonic \cite{Photonic_2019}, polaronic \cite{Cerda2013} and atomic systems \cite{Atomic_2019}.

  Gamayun \textit{et. al} demostrated the absence of a gap for a Kek-Y distortion and deduced the low energy Hamiltonian for Kekul\'e distortions by using a first-neighbor tight-binding Hamiltonian \cite{Gamayun}. They show how the two Dirac cones merge at the center of the Brillouin zone, producing either a gap (Kek-O) or the superposition of two cones with different Fermi velocities (Kek-Y)\cite{Gamayun}. 
 
 Several works have been made using such  first-neighbor tight-binding Hamiltonian, for example to study uniaxial strain\cite{Elias_2019,Eom2020Direct} and  the electronic transport properties \cite{KuboKek,Elias2020,Barrier-Kek,kek-device}.

However, its is known that second-neighbors interactions in graphene are very important \cite{Castro_Neto,Botello}. They are
fundamental to explain the electronic properties at graphene surface as in graphene nanoribbons  \cite{Castro_Neto}. This leads to the natural question of what are the effects of second neighbors interactions in a Kekulké patterns. Although Density Functional Calculations   already contains such effects, due to the involved energies and the low resolution of the mesh calculations near the Dirac cones, its is difficult to assert
a detailed picture of the energy dispersion. In that sense, a tight-binding calculation can be very useful. Here we tackle this question by producing a low-energy Hamiltonian for a Kekulé patterned graphene which includes second-neighbor interactions. The resulting model is validated through a comparison with the numerical calculations.   

The paper is organized as follows. In Sec. \ref{sec:Hamiltonian} we introduce the Hamiltonian for a honeycomb lattice with a Kek-Y distortion up to next-nearest neighbors, and in Sec. \ref{sec:lowenergy} we calculate an effective four band low-energy Hamiltonian, finally in \ref{sec:conclusions} we present our conclusions and remarks.

\section*{First and second neighbor Kekul\'e-Y graphene Hamiltonian}\label{sec:Hamiltonian}
\begin{figure}%
    \centering
    {\includegraphics[scale=0.5]{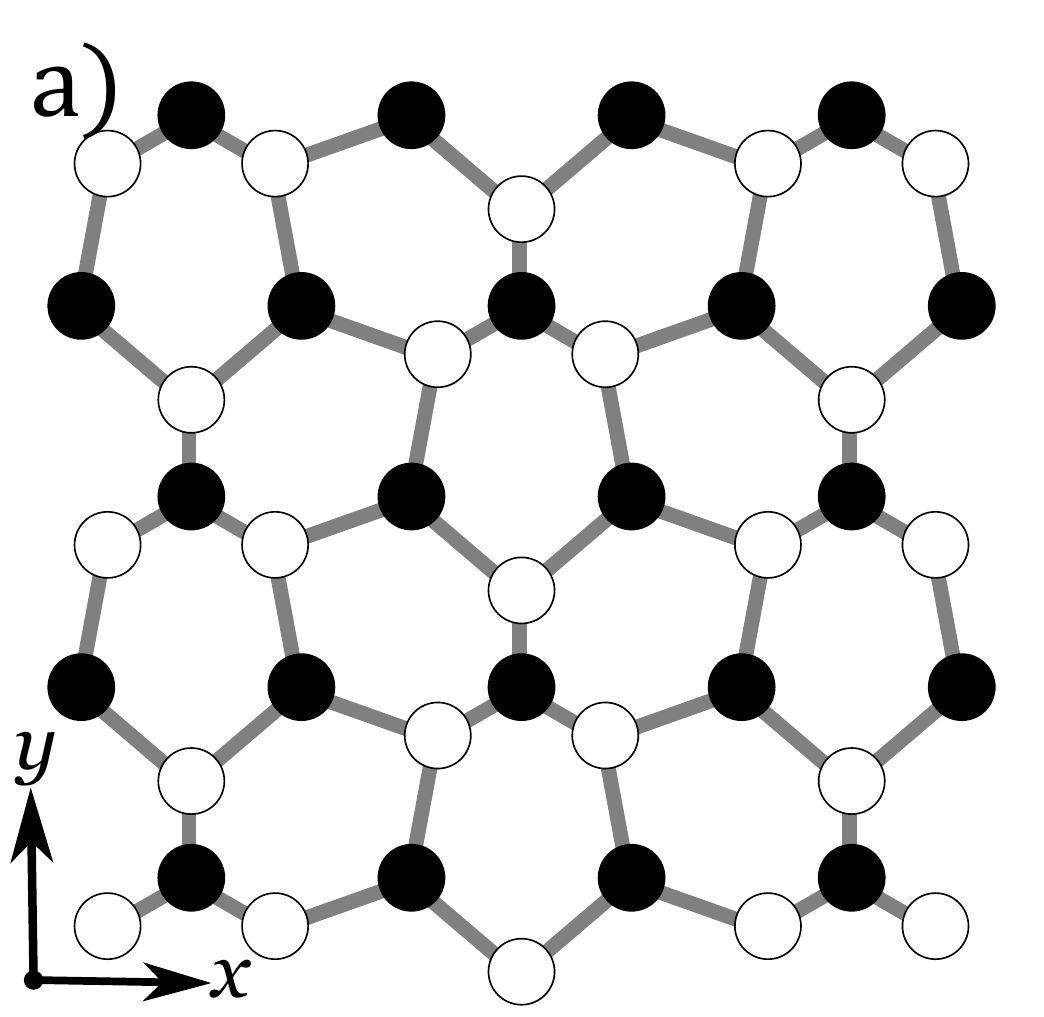}}%
    {\includegraphics[scale=0.5]{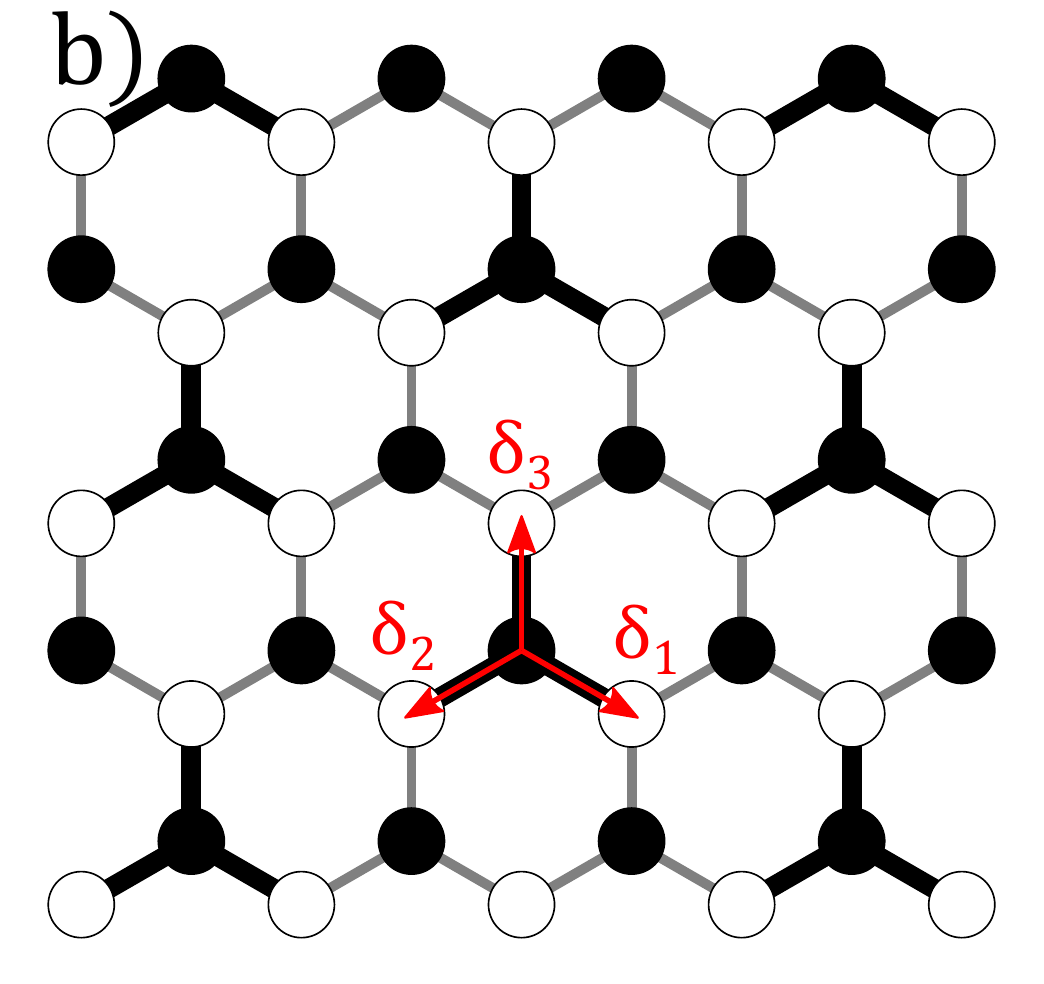}}%
    {\includegraphics[scale=0.5]{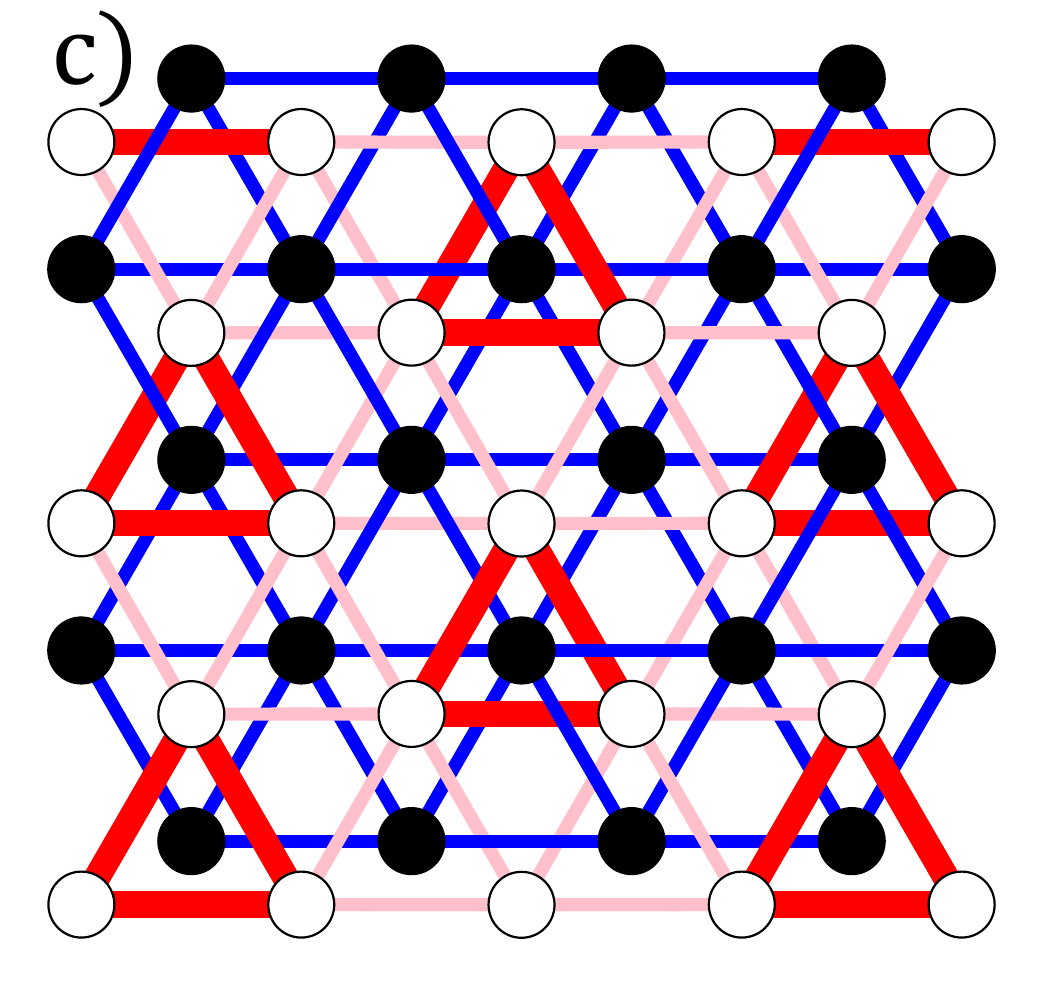}}%
    \caption{Kek-Y distorted graphene lattice. The A and B sublattices are denoted by the black and white circles respectively. a) Deformed lattice, not in scale for illustration purposes. b) Nearest neighbors hoppings. The bold black lines indicate a stronger bond than the gray ones. c) Next-nearest neighbors interactions scheme. They form two independent triangular lattices for each sublattice, here shown in blue for the one corresponding to the A sublattice. These A-A bonds remain unaltered by the Kek-Y pattern. For the B sublattice, a B-B bond texture is induced in such a way that the bonds near the Y deformation (bold red lines) are stronger than the others (pink lines).
    }%
    \label{Fig:syst}%
\end{figure}
We can consider the Kek-Y bond modulation as a periodic strain which reduces the distance between one third of the atoms in one sublattice with its three nearest neighbors as shown in Fig. \ref{Fig:syst}a. Thus the hopping integral gets modified by the change in the interatomic distances accordingly with strain theory \cite{naumis2017electronic}. However, this would also mean a change in the next-nearest neighbors hoppings. In Fig. \ref{Fig:syst}c we illustrate this point, by showing in red (pink)  the stronger (weaker) bonds for the B sublattice. In blue we show the hoppings between the A sublattice atoms, which remain unaltered.

Assuming the Gruneissen parameter to be equal for both first and second neighbors we can consider that the bond changes with the same proportionality. Thus the Hamiltonian for graphene with Kek-Y distortion considering hopping up to next-nearest neighbors is,
\begin{equation}
    H=-\sum_{\bm r} \sum_{j=1}^{3} t^{(0)}_{\bm r} a_{\bm r}^{\dagger}b_{\bm{r+\delta_j}}+h.c.+\sum_{\bm r} \sum_{m\neq n} t_2 a_{\bm r}^\dagger a_{\bm{r+\delta_m-\delta_n}}+\sum_{\bm r} \sum_{m\neq n} t^{(2)}_{\bm r} b_{\bm r+\delta_m}^\dagger b_{\bm{r+\delta_n}},
\end{equation}
where $\bm r$ is the position vector that runs over the atomic positions of sites in sublattice A and its given by $\bm r = n_1 \bm a_1 + n_2 \bm a_2 $ where $n_1$, $n_2$ are integers, $\bm a_1=a\left ( -\frac{\sqrt{3}}{2},\frac{3}{2} \right)$, $\bm a_2=a\left ( \frac{\sqrt{3}}{2},\frac{3}{2} \right)$ are the lattice vectors and $a$ is the distance between carbon atoms $1.42$ \AA. The vectors $\delta_1=a\left (\frac{\sqrt{3}}{2},-\frac{1}{2} \right)$, $\delta_2=a\left (-\frac{\sqrt{3}}{2},-\frac{1}{2} \right)$ and $\delta_3=a\left (0,1 \right)$ go from a site in the A sublattice towards its three nearest neighbors in the B sublattice as shown in Fig. \ref{Fig:syst}. The space dependent hopping parameters $t_{\bm r}^{(0)}$ and $t_{\bm r}^{(2)}$ between nearest neighbors and next-nearest neighbors respectively, are periodically modulated as
\begin{equation}
    t_{\bm r}^{(0)}/t_0=t_{\bm r}^{(2)}/t_2=1+2 \Delta \text{cos}(\bm G \cdot \bm r),
\end{equation}
here $t_0=2.8$ eV is the hopping parameter for first neighbors in pristine graphene, and $t_2$ for second neighbors which will be taken as $t_2=0.1t_0$ unless otherwise indicated. $\Delta$ is the Kekulé coupling amplitude and $\bm G=\frac{4 \pi}{9}\sqrt{3}(1,0)$ is the Kekulé wave vector.
The Fourier transform of our tight-binding Hamiltonian is then,
\begin{equation}
    \begin{split}
    H(\bm k)=&-\epsilon(\bm k)a^{\dagger}_{\bm k} b_{\bm k}-\Delta \epsilon(\bm{k+G})a_{\bm{k+G}}^{\dagger}b_{\bm k}-\Delta \epsilon(\bm{k-G})a_{\bm{k-G}}^{\dagger}b_{\bm k} + h.c.\\
    &+f(\bm k) a_{\bm k}^{\dagger} a_{\bm k}+f(\bm k) b_{\bm k}^{\dagger} b_{\bm k}+\Delta f^+(\bm k) b_{\bm {k+G}}^{\dagger}b_{\bm k}+\Delta f^-(\bm k) b_{\bm {k-G}}^{\dagger}b_{\bm k},
    \end{split}
\end{equation}
where
\begin{subequations}
    \begin{equation}
        \epsilon(k)=t_0 \sum_{j=1}^3 e^{i \bm{k \cdot \delta_j}},
    \end{equation}
    \begin{equation}
        f(k)=t_2 \sum_{m\neq n} e^{i \bm{k \cdot (\delta_m-\delta_n)}},
    \end{equation}
    \begin{equation}
        f^{\pm}(k)=t_2 \sum_{m\neq n} e^{i \bm{k \cdot (\delta_m-\delta_n)}}e^{\mp i\bm{G \cdot \delta_n}},
    \end{equation}
\end{subequations}
here $\epsilon(\bm k)$ and $f(\bm k)$ are the dispersion relations for a honeycomb and a triangular lattice respectively. Some properties of  $f^{\pm}(\bm k)$ are:
\begin{equation}
    f^{\pm^*}(\bm k \pm \bm G)=f^{\mp}(\bm k \mp \bm G), \quad f^{\mp^*}(\bm k \pm \bm G)=f^{\pm}(\bm k).
\end{equation}
By defining the column vector $c_{\bm{k}}=(a_{\bm{k}},b_{\bm{k}},a_{\bm{k+G}},b_{\bm{k+G}},a_{\bm{k-G}},b_{\bm{k-G}})$ we can rewrite the Hamiltonian as a 6 x 6 matrix:
\begin{subequations}
\begin{equation}\label{eq:Hdek}
    H(\bm k)=c_{\bm k}^{\dagger}
    \begin{pmatrix}
    H_{\Gamma} & T \\
    T^{\dagger} & H_{G}
    \end{pmatrix}
    c_{\bm k},
\end{equation}
made from the original $\Gamma$ point graphene Hamiltonian, 
\begin{equation}
    H_{\Gamma}=
    \begin{pmatrix}
    f(\bm k) & -\epsilon(\bm k) \\
    -\epsilon(\bm k) & f(\bm k)\\
    \end{pmatrix},
\end{equation}
the $\bm{G}$ and $\bm{-G}$ points Hamiltonian,  
\begin{equation}
    H_{G}=
    \begin{pmatrix}
    f(\bm{k+G}) & -\epsilon(\bm{k+G}) & 0 & -\Delta \epsilon(\bm{k-G})\\
    -\epsilon^*(\bm{k+G}) & f(\bm{k+G}) & -\Delta \epsilon^*(\bm{k+G}) & \Delta f^-(\bm{k-G})\\
    0 & -\Delta \epsilon(\bm{k+G}) & f(\bm{k-G}) & -\epsilon(\bm{k-G})\\
    -\Delta \epsilon^*(\bm{k-G}) & \Delta f^+(\bm{k+G}) & -\epsilon^*(\bm{k-G}) & f(\bm{k-G})\\
    \end{pmatrix},
\end{equation}
and the interaction between them,
\begin{equation}
    T=
    \begin{pmatrix}
    0 & -\Delta \epsilon(\bm{k+G}) & 0 & -\Delta \epsilon(\bm{k-G})\\
    - \Delta \epsilon^*(\bm k) & \Delta f^-(\bm{k+G}) & - \Delta \epsilon^*(\bm k) & \Delta f^+(\bm{k-G})
    \end{pmatrix}.
\end{equation}
\end{subequations}

\begin{figure}[!htbb]
\begin{center}
\includegraphics[scale=0.5]{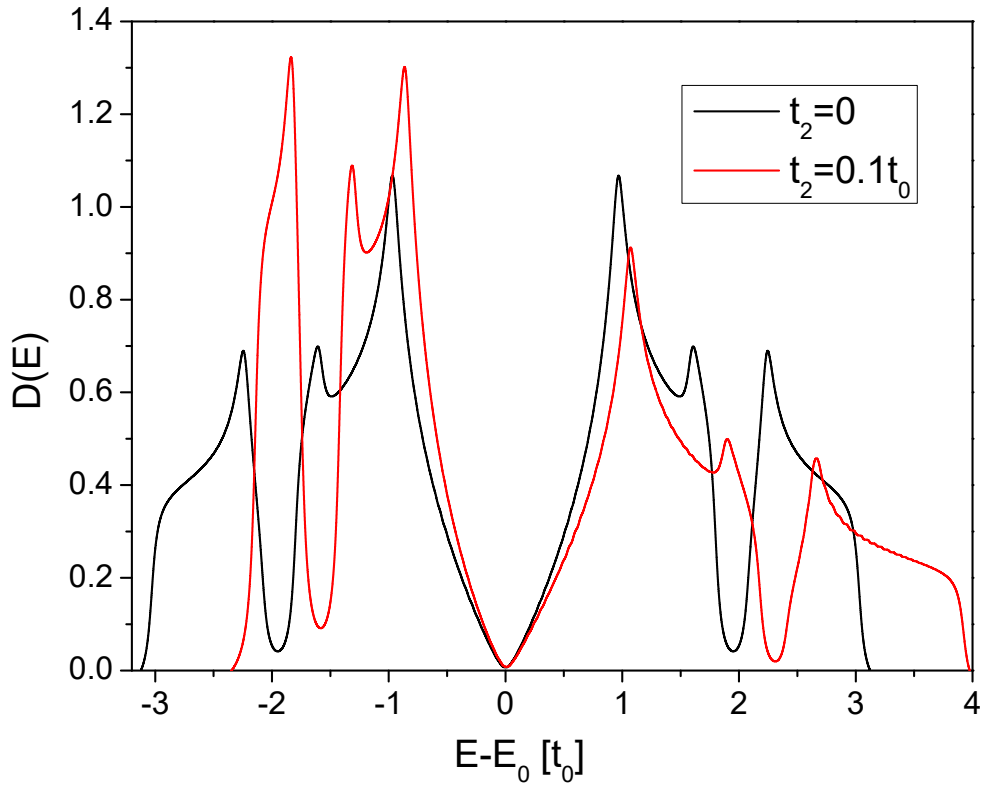}
\end{center}
\caption {Comparison of the DOS for Kekul\'e patterned graphene with and without the second-neighbor interaction, where the zero energy corresponds to the Fermi energy. Notice the electron-hole asymmetry which is specially clear for the band widths.} 
\label{Fig:DOS}
\end{figure}

In Fig. \ref{Fig:DOS} we present the density of states (DOS) obtained from a numerical diagonalization of Eq. (\ref{eq:Hdek}). As expected, the main effect is the breaking of the electron-hole symmetry reflected in changes of the band widths. Also, around the Fermi energy there are changes that we explore in the following section, as we will develop a low-energy approximation and compare it with the numerical diagonalization of the Hamiltonian given in (\ref{eq:Hdek}).

\section*{Low-Energy Hamiltonian}\label{sec:lowenergy}
Let us now build a low energy Hamiltonian starting with the full $6\times 6$ Hamiltonian given by Eq. (\ref{eq:Hdek}). Since we are interested in the low energy bands, we expand up to first order in $\bm k$ the functions that appear in Eq. (\ref{eq:Hdek}). We obtain the following results,
\begin{subequations}
    \begin{equation}
        \epsilon(\bm{k}) \approx 3 t_0, \quad \epsilon(\bm{k\pm G}) \approx \frac{3}{2} t_0 (\mp k_x + i k_y),
    \end{equation}
    \begin{equation}
        f(\bm{k}) \approx 6t_2, \quad f(\bm{k\pm G}) \approx -3t_2,
    \end{equation}
    \begin{equation}
        f^{\pm}(\bm{k\pm G}) \approx 0, \quad f^{\pm}(\bm{k\mp G}) \approx \frac{9}{2} t_2 (\pm k_x+i k_y), \quad f^{\pm}(\bm{k}) \approx \frac{9}{2} t_2 (\mp k_x - i k_y).
    \end{equation}
\end{subequations}
Using the previous approximations, the linearized components of the  Hamiltonian Eq. (\ref{eq:Hdek}) are given by,
\begin{subequations}
\begin{equation}
    H_{\Gamma} \approx
    \begin{pmatrix}
    6 t_2 & -3 t_0 \\
    -3 t_0 & 6t_2\\
    \end{pmatrix},
\end{equation}
\begin{equation}
    H_{G} \approx
    \begin{pmatrix}
   -3t_2 & v_F \hbar (k_x-ik_y) & 0 & -\Delta v_F \hbar(k_x+ik_y)\\
   
    v_F \hbar (k_x+ik_y) & -3t_2 & \Delta v_F \hbar (k_x+ik_y )& 0 \\\
    
    0 & \Delta v_F \hbar (k_x-ik_y) & -3t_2 & -v_F \hbar (k_x+ik_y)\\
    
    -\Delta v_F \hbar (k_x-ik_y)) & 0 & -v_F \hbar (k_x-ik_y) & -3t2\\
    \end{pmatrix},
\end{equation}
\begin{equation}
    T \approx
    \begin{pmatrix}
    0 & \Delta v_F \hbar (k_x-ik_y) & 0 & -\Delta v_F \hbar (k_x+ik_y)\\
    -3\Delta t_0 & -\Delta v_2 \hbar (k_x-ik_y) & -3\Delta t_0 & \Delta v_2 \hbar (k_x+ik_y)
    \end{pmatrix},
\end{equation}
\end{subequations}
where we defined two velocities, one is the usual Fermi velocity in pristine graphene, 
\begin{equation}
v_F=\frac{3a t_0}{2 \hbar}
\end{equation}
and the other is due to second neighbors, 
\begin{equation}v_2=\frac{9a t_2}{2 \hbar}=3\left(\frac{t_2}{t_0}\right)v_F.
\end{equation}
As we are interested in the spectrum at low energies,  the relevant part is the one associated with the valleys $\bm K$ and $\bm{K'}$ which corresponds to the block matrix $H_G$, and then we can add the effects from higher energy bands in the $\bm{\Gamma}$ point as a perturbation. We have verified that this is an essential step in order to recover the spectrum obtained from a direct diagonalization. We can obtain the effective Hamiltonian by projecting $H_{\Gamma}$ into the subspace of $H_G$. To do this, consider the Schrodinger equation applied to Eq. (\ref{eq:Hdek}),
\begin{equation}
  H_{\Gamma}\Psi_{\bm{\Gamma}}+T\Psi_{\bm{G}}=E\Psi_{\Gamma}
\end{equation}
and
 \begin{equation}
 T^{\dagger}\Psi_{\bm{\Gamma}}+H_{\bm{G}}\Psi_{\bm{G}}=E\Psi_{\bm{G}}
\end{equation}
where $\Psi_{\Gamma}$ and $\Psi_{\bm{G}}$ are the components of the solution $\Psi=(\Psi_{\Gamma},\Psi_{\bm{G}})$ on each subspace. 

From the first equation we can obtain $\Psi_{\Gamma}$ and
use it on the second to obtain an effective Hamiltonian for the $\Psi_{\bm{G}}$ component resulting in an effective Hamiltonian, 
\begin{equation}
    H_{Eff}=H_G+T^{\dagger}(E \mathds{1}-H_{\Gamma})^{-1}T
\end{equation}
This Hamiltonian is exact but needs a self-consistent procedure to find $E$. However, if we expand the term  $(E\mathds{1}-H_{\Gamma})^{-1}$ and keep the first order term. we can make the approximation $E\approx E_0=-3t_2$ which is the original energy dispersion in the $\Gamma$ point. Now we write the Dirac-like equation for this system,
\begin{subequations}
\begin{equation}
    \mathcal{H}
    \begin{pmatrix}
    \Psi_{\bm K'}\\
    \Psi_{\bm K}
    \end{pmatrix}
    =E
    \begin{pmatrix}
    \Psi_{\bm K'}\\
    \Psi_{\bm K}
    \end{pmatrix} ,   
\end{equation}
\begin{equation}
    \Psi_{\bm K'}=
    \begin{pmatrix}
    -\psi_{B,\bm K'}\\
    \psi_{A,\bm K'}
    \end{pmatrix},
    \quad
    \Psi_{\bm K}=
    \begin{pmatrix}
    \psi_{A,\bm K}\\
    \psi_{K,\bm K}
    \end{pmatrix},
\end{equation}
where the explicit form of the low-energy Hamiltonian is finally given by,
\begin{equation*}
    \mathcal{H}=
    \begin{pmatrix}
    E_0 & v_F (1-\Delta^2)(p_x-ip_y) & v_F \Delta (1-\Delta) (p_x-ip_y) & 0\\
    v_F (1-\Delta^2)(p_x+ip_y) & E_0+\mu & \mu & v_F \Delta (1-\Delta) (p_x-ip_y)\\
    v_F \Delta(1-\Delta^2)(p_x+ip_y) & \mu & E_0+\mu & v_F (1-\Delta^2)(p_x-ip_y)\\
    0 & v_F \Delta (1-\Delta^2)(p_x+ip_y) & v_F (1-\Delta^2)(p_x+ip_y) & E_0
    \end{pmatrix},
\end{equation*}
which can be compactly written as,
\begin{equation}
\begin{split}
    \mathcal{H}=& E_0 \bm \sigma_0 \otimes \bm \tau_0 + v_f (1-\Delta^2) (\bm p \cdot \bm \sigma) \otimes \bm \tau_0 + v_f \Delta (1-\Delta) \bm \sigma_0 \otimes (\bm p \cdot \bm \tau)\\
    & + \frac{\mu}{2}(\bm \sigma_0 \otimes \bm \tau_0 + \bm \sigma_x \otimes \bm \tau_x +\bm \sigma_y \otimes \bm \tau_y - \bm \sigma_z \otimes \bm \tau_z),
\end{split}
\end{equation}

\end{subequations}
with $\mu$ defined as,
\begin{equation}
    \mu=\frac{9 \Delta^2 t_0^2 t_2}{t_0^2-9t_2^2}.
\end{equation}

The four low-energy bands are
\begin{subequations}
\begin{equation}
    E_{D}^{\pm}=E_0 \pm v_D |\bm p|,
\end{equation}
\begin{equation}
    E_{M}^{\pm}=E_0+\mu \pm \sqrt{v_M^2 \bm p^2 +\mu^2},
\end{equation}
\label{Eq:DispLE}
\end{subequations}
were we defined $v_D=v_F(1-\Delta)$ and $v_M=v_F(1-\Delta)(1+2\Delta)$.

\begin{figure}[!htbb]
\begin{center}
\includegraphics[scale=0.5]{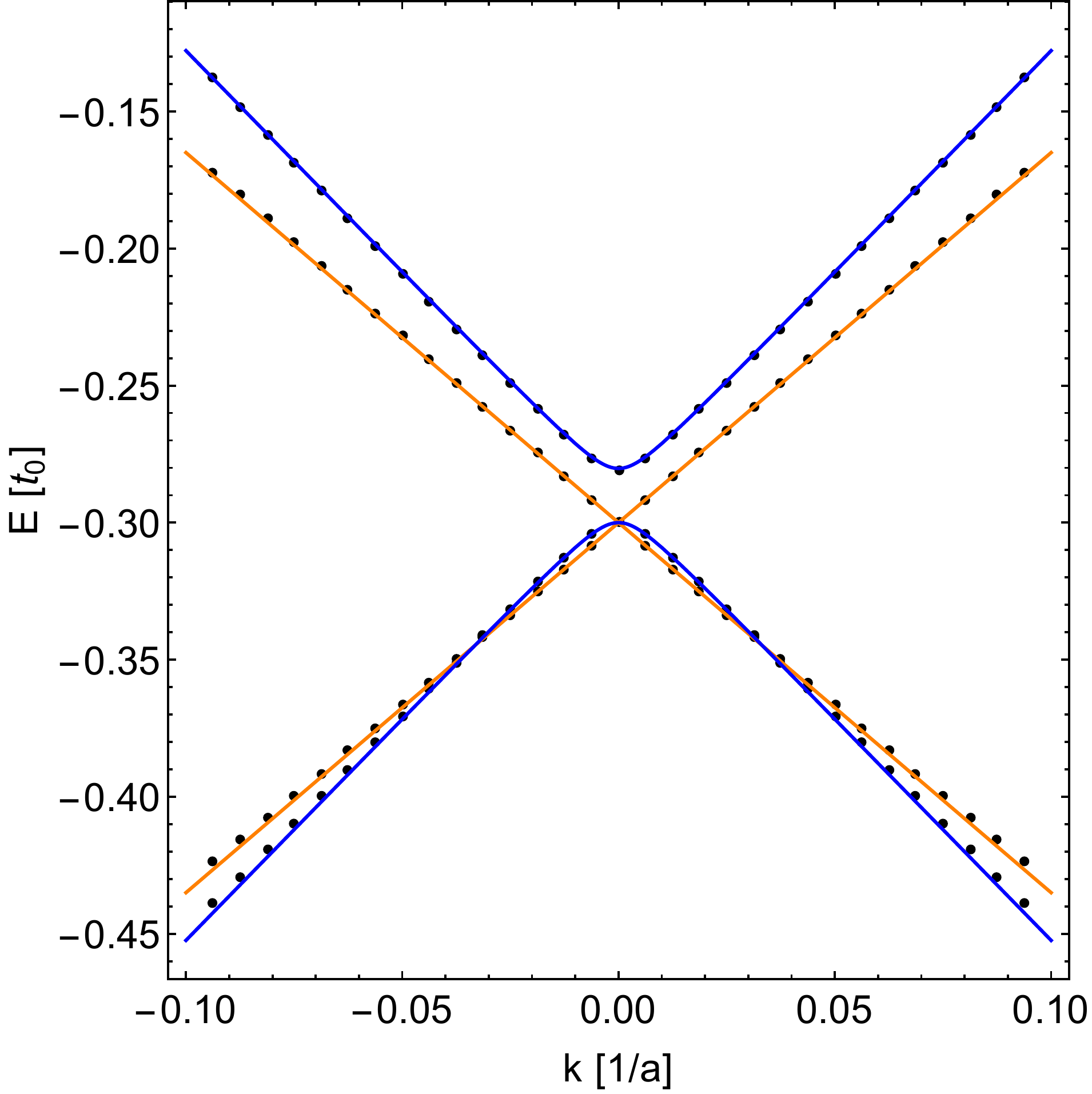}
\end{center}
\caption {Low-energy dispersion around the $\Gamma$ point with $\Delta=0.1$ and $t_2/t_0=0.1 $. The solid lines indicate our analytic results for both the conic (orange) and the bands with an effective mass (blue). The numerical tight-binding calculations obtained by a direct diagonalization of the Hamiltonian are represented by the dots.} 
\label{Fig:Disp}
\end{figure}

Therefore, there is a mix of two-flavor Fermion gases. One with and effective Dirac Hamiltonian,
\begin{equation}
    H_{D}=E_0\bm{\sigma}_0+v_{D}\bm{p} \cdot \bm{\sigma},
\end{equation}
and the other,
\begin{equation}
    H_M= (E_0+\mu)\bm{\sigma}_0+v_{M}\bm{p} \cdot \bm{\sigma} + \mu \bm{\sigma}_z.
\end{equation}
In Fig. \ref{Fig:Disp} we compare our results from Eq. (\ref{Eq:DispLE}) with the numerical diagonalization of the tight-binding Hamiltonian given by Eq. (\ref{eq:Hdek}). First we notice an excellent agreement within this regime. Without second neighbors interaction, the Kek-Y bond texture couples both valleys in the $\Gamma$ point, resulting in two concentric cones with different velocities \cite{Gamayun}. Turning on the interaction gives rise to two main effects, a set of bands gains an effective mass and a shift in energy. This last effect results in a  particle-hole symmetry breaking, lifting the degeneracy of the conduction bands at the Dirac point, therefore only three bands intersect. This structure is akin to a “spin-one Dirac cone", expected for a honeycomb lattices with a distinction between one third of the atoms in one sublattice \cite{single_valley,pseudospin_one}. We can see that the effect of adding next-nearest-neighbors interaction is equivalent to that of an on-site potential $\mu$ on the atom at which the Y deformation is centered \cite{Gamayun}.

\begin{figure}[!htbb]
\begin{center}
\includegraphics[scale=0.4]{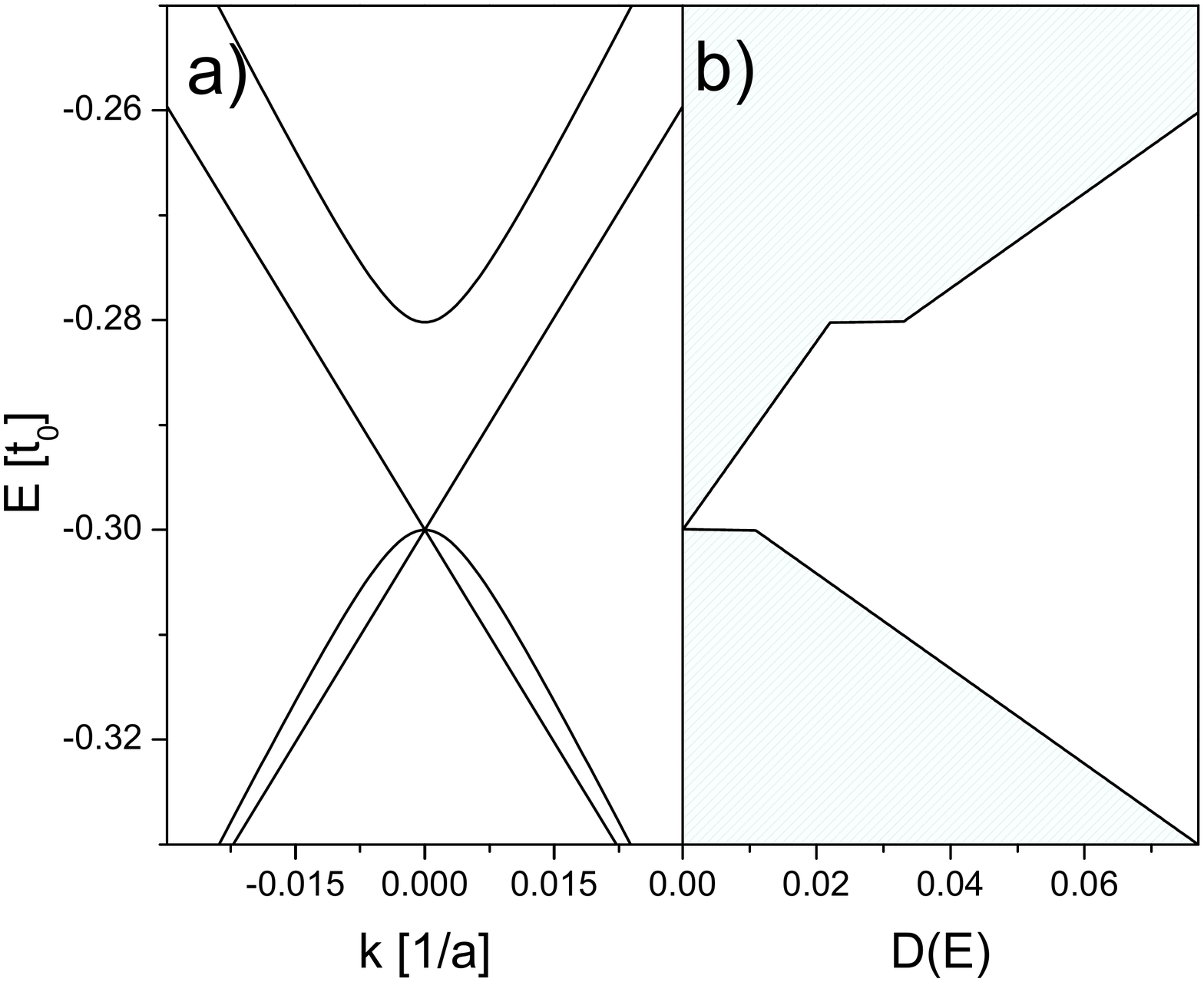}
\end{center}
\caption {a) A zoom of the energy dispersion around the Fermi energy of Kekul\'e patterned graphene including up to second-neighbor interactions  obtained from the low-energy approximation. b) The corresponding density of states showing how the cone with an effective mass produces two jumps in the DOS.} 
\label{Fig:LEDOS}
\end{figure}

From Eq. (\ref{Eq:DispLE}) we can easily calculate the density of states per unit cell. Considering spin degeneracy,  it is given by,
 \begin{equation}
    D(E)=\frac{A}{\pi \hbar^2} \left [\frac{|E-E_0|}{v_D^2}+\frac{E-(E_0+\mu)}{v_M^2} \Theta(E-E_0-2\mu)+\frac{(E_0+\mu)-E}{v_M^2} \Theta(E_0-E)\right],
 \end{equation}
 where $A=9 \sqrt{3}a^2/2$ is the unit cell area and $\Theta(E)$ is the Heaviside function. Although the density of states retains its linear behavior around the Dirac point, the massive bands produce a discontinuity shown in Fig. \ref{Fig:LEDOS}.

\begin{figure}[!htbb]
\begin{center}
\includegraphics[scale=0.5]{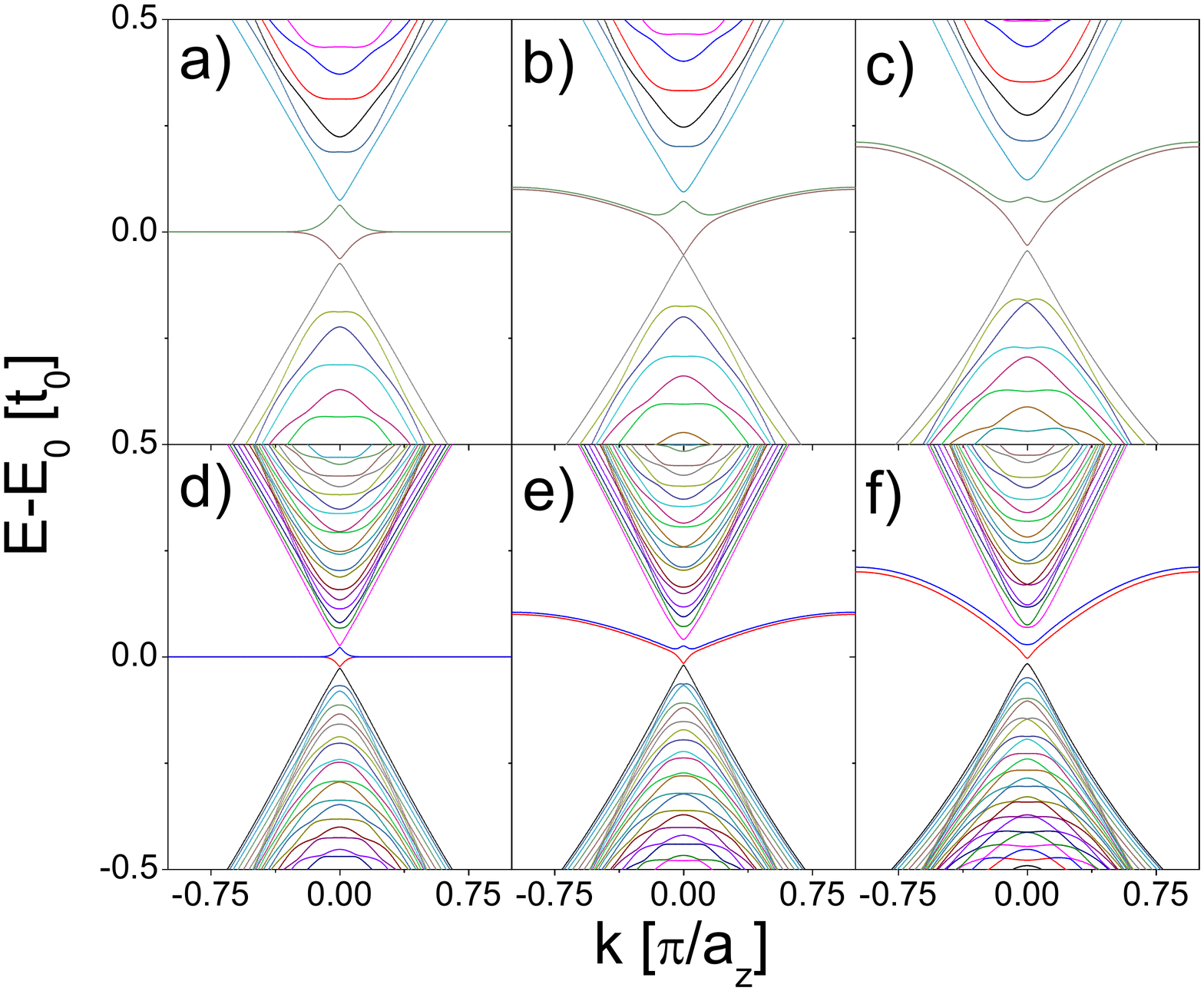}
\end{center}
\caption {Dispersion for zigzag graphene nanoribbons with Kekulé-Y bond texture and hoppings up to next nearest neighbors. The amplitude of the bond texture is $\Delta=0.1$ for all figures. a), b) and c) show the dispersion with $t_2/t_0=0,0.1$ and $0.2$ respectively for a ribbon with $W=4.54$ nm, analogously d), e) and f) for a ribbon with $W=13.06$ nm.}
\label{Fig:Ribbons}
\end{figure}

Second neighbors hoppings are particularly important for graphene nanoribbons (GNR). We calculated numerically the band structure for zigzag edged GNR. In Fig. \ref{Fig:Ribbons} our results are shown for different values of $t_2$ and width $W$. Due to the change in the periodicity produced by the Kekulé texture, the unit cell size $a_z$ is three times bigger, thus $a_z=3 \sqrt{3} a$.
We can see that edge states become dispersive, which is a well known effect of second neighbors interaction \cite{Castro_Neto}, however the combination with the Kek-Y bond texture results in an hybridization of both edge states. The velocity induced in the edge states may indirectly close the small gap predicted for Kek-Y zigzag GNR \cite{Elias2020}.

\section{Conclusions}\label{sec:conclusions}

The effects of second-neighbor interactions in
Kekul\'e patterned graphene were studied starting from a tight-binding Hamiltonian. From there, a low-energy effective Hamiltonian was derived using a projection technique. This Hamiltonian was validated thorough a comparison with a numerical calculation obtained from the diagonalization of the full tight-binding Hamiltonian. We found that beyond the expected electron-hole symmetry breaking, the main effect of the second-neighbor interaction is that in one of the Dirac cones, the electron becomes massive when compared with the calculation  made considering only first-neighbour interaction. As a result, the density of states near the Fermi energy contains  a jump in the otherwise linear behavior. Finally, we considered the effects of second-neighbor interactions in Kekul\'e patterned graphene nanoribbons, as it is known that such effects are essential to reproduce a minimally realistic behavior at the edges. As expected, the same mass effect is seen in the nanoribbons and in fact is amplified as the width is decreased.   

Thus, we expect that such second-neighbor effects to be important in the electronic and optical properties of Kekul\'e bond textures.\\

We thank UNAM DGAPA project IN102620 and CONACYT project 1564464. E. Andrade thanks an schoolarship from CONACyT.

\bibliographystyle{unsrt}
\bibliography{KekYNNNH}

\end{document}